\documentclass[twocolumn,amsmath,amssymb,prb]{revtex4}
\usepackage{graphicx}

\begin{document}

\title{Peculiarities of dynamics of Dirac fermions associated with zero-mass lines}
\author{Timur Tudorovskiy}
\author{Mikhail I. Katsnelson}
\affiliation{Institute for Molecules and Materials,\\
Radboud University of Nijmegen,\\
Heyendaalseweg 135, 6525AJ Nijmegen, The Netherlands}

\date{\today}

\begin{abstract}
Zero-mass lines result in appearance of linear dispersion modes for Dirac fermions. These modes play an important role in various physical systems.
However, a Dirac fermion may not precisely follow a single zero-mass line, due to either tunneling between different lines or centrifugal
forces. Being shifted from a zero-mass line the Dirac fermion acquires mass which can substantially influence its expected ``massless''
behavior. In the paper we calculate the energy gap caused by the tunneling between two zero-mass lines and show that its opening leads
to the delocalization of linear dispersion modes. The adiabatic bending of a zero-mass line gives rise to geometric phases. These are the Berry phase, locally associated with a curvature, and a new phase resulting from the mass square asymmetry in the vicinity of a zero-mass line.
\end{abstract}


\maketitle

\section{Introduction}

Zero energy states are well known for the one-dimensional eigenvalue-problem for the Dirac particle with a spatially dependent mass. Such a problem naturally arises in various contexts, both in condensed matter and in high-energy physics \cite{kri87}. These are super-symmetric quantum mechanics \cite{wit81}, fractional charge \cite{jac76} and solitons in polyacetylene \cite{su79}. For the monotonous mass distribution the wavefunction corresponding to the zero energy state is localized in the vicinity of the point were the mass vanishes.

Let us consider the two-dimensional case. Assuming that the mass depends on a single variable only, say $y$, we come back to the one-dimensional problem. However, in two dimensions the motion is allowed not only along the $y$-axis, but also along the perpendicular to it $x$-axis. If the mass vanishes at the point $y_0$, the line $(x,y_0)$ is the zero-mass line (ZML). Our two-dimensional problem reduces to the one-dimensional problem completely if we assume that the particle does not move along ZML. For non-zero values of the momentum $p_x$ along ZML the energy of the particle is given by the linear relation, $E=\pm p_x$, i.e. the one-dimensional zero-energy state becomes a linear dispersion mode (LDM) in two dimensions.

LDMs naturally appear as edge states for inverted band semiconductors \cite{volk86} as well as for a certain model of the quantum Hall effect\cite{lud94}. Though single-particle LDMs in quantum Hall regime might not be a sufficient description for conventional two-dimensional electron gas\cite{chk92}, this description seems to be adequate for the case of narrow graphene ribbons \cite{het12}. Another recent examples where ZML are essentially involved are given by topologically protected edge states in CdTe/HgTe/CdTe topological insulators \cite{koe07,M09,HK10,bue11,qi11}, graphene on boron nitride \cite{sac11,zar12}, and chemically functionalized graphene \cite{eli09,bal10,rah10} where for an inhomogeneous functionalization the mass term can in general change its sign. LDMs arise in gapped bilayer graphene \cite{mar08,li11,zar11} and chiral p-wave superconductors \cite{bar10}. Changing a width of CdTe/HgTe/CdTe quantum well or applying a gate voltage one can also create ZML in bulk.
Though our consideration will be formally applicable in all these cases, for the sake of definiteness we assume in this paper that LDMs relate to edge states in two-dimensional topological insulators.

An effective dynamics of charge carriers in topological insulators \cite{M09,HK10,qi11} is governed by the Dirac Hamiltonian with a spatially dependent mass term. This term vanishes along ZML giving rise to LDM. This mode lies in a gap for bulk states. Along a single straight ZML the linear dispersion mode propagates only in one direction. This unidirectional edge mode is very similar to unidirectional edge states in conventional semiconductors placed in the high magnetic field. It is well known, that these states support the quantum Hall current \cite{hal82}. The existence of unidirectional channels in topological insulators in zero magnetic field is referred to, by analogy with the quantum Hall effect, as quantum spin Hall effect \cite{KM05b,ber06,koe07}.

\begin{figure}
\begin{center}
\includegraphics[width=7cm]{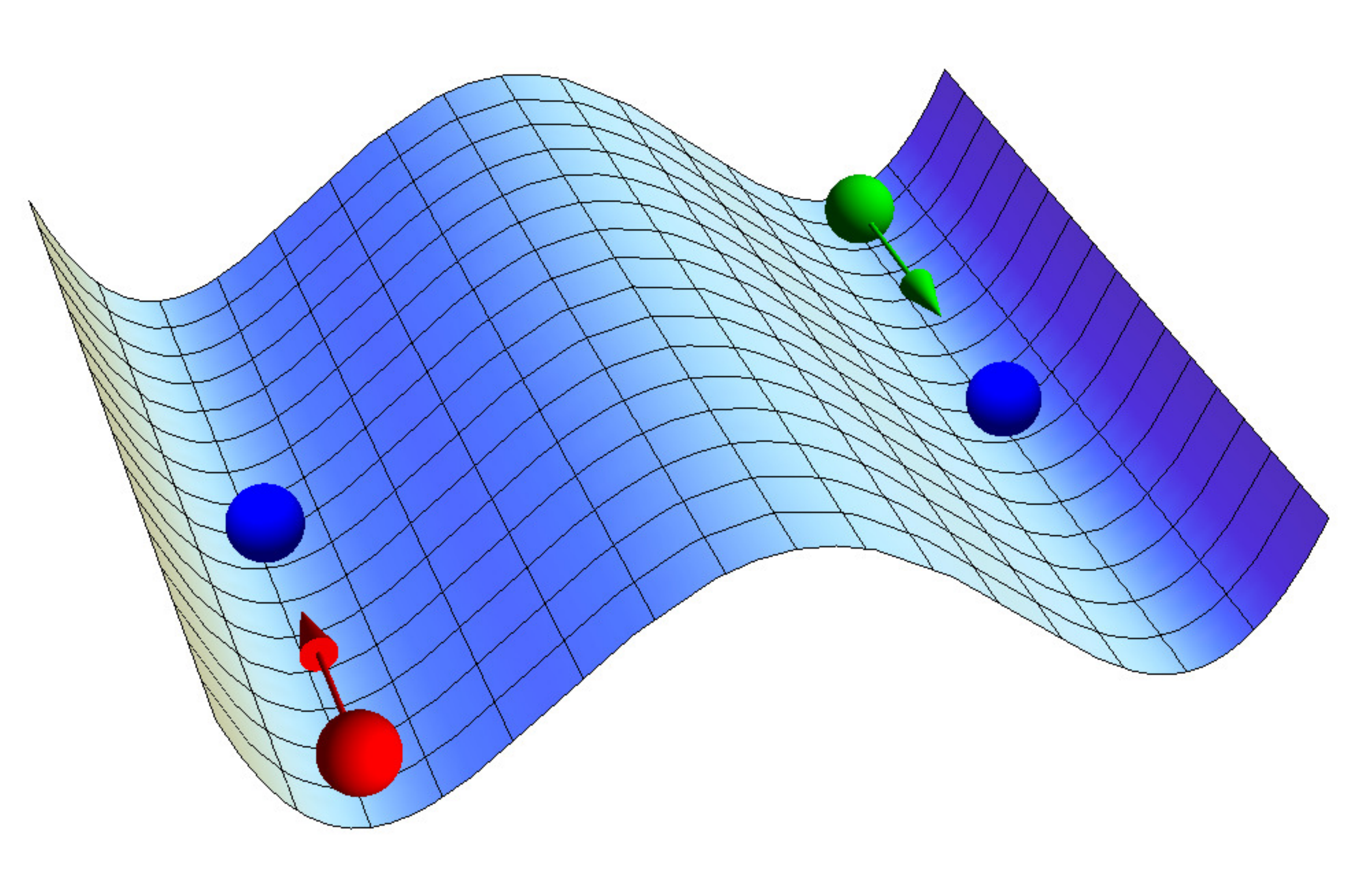}
 \caption{\label{fig::topo}Particles corresponding LDM on the mass square surface.
 Red and green spheres illustrate edge states with large longitudinal momenta. They are localized
 at a single zero-mass line. Blue spheres illustrate states with small longitudinal momenta. Such states
 are localized at both ZMLs simultaneously.}
\end{center}
\end{figure}

From the physical point of view a mass square landscape in a topological insulator forms a waveguide around ZML (see Fig.~\ref{fig::topo}). LDM in such a waveguide corresponds to zero transversal momentum, i.e. to the rest in the direction transversal to ZML. This stands in stark contrast to the conventional Schr\"odinger particle in a soft-walls waveguide, where the lowest transversal energy is positive.

LDMs in topological insulators are known to be topologically protected against scattering by non-magnetic
impurities \cite{qi11}. Topological protection results from the spatial separation of states traveling in different directions,
thus the backscattering should be attenuated by the probability to tunnel through the bulk of a sample.
This peculiarity is very similar to the behavior of conducting states in quantum Hall effect. In ideal system at zero temperature
it is the tunneling between edge states which determines an accuracy of quantum Hall plateaus.

The assumption about the spatial separation of states traveling in different directions does not take into account
peculiarities of quantum tunneling. Indeed, let us consider a ribbon with two parallel ZML at its opposite
edges and assume that mass does not depend on the variable along the line.
One can expect that the crossing of straight lines corresponding to a linear dispersion turns into the avoided crossing.
Obviously, this effect does not depend on the symmetry: a transition from a crossing to an avoided crossing caused by tunneling is generic.

Due to the gap opening the LDM cone turns into two branches, almost linear at large momenta. Let us consider an upper branch. For large negative
momenta the wavefunction is localized at one edge of the sample and for large positive momenta it is localized at
the other edge. Since the wavefunction smoothly depends on the longitudinal momentum, we should conclude that at a certain longitudinal momentum amplitudes of the wavefunction are comparable at both ZMLs. At this point states traveling in different directions are not spatially separated and the topological protection may break (see Fig.~\ref{fig::topo}). In Section \ref{zz} we illustrate this effect by an example of a symmetric mass distribution.

The tunneling between ZMLs has been studied experimentally \cite{kan00}. In this work the differential conductance measurements in the integer quantum Hall regime between two parallel edge channels were reported. Theoretically\cite{str09} one considered the tunneling between Quantum Hall edge states via the Landau-Zener-like mechanism: it was assumed that the main contribution comes from the vicinity of a point were edge channels are the most close to each other.

Topological protection of a single ZML should lead to a substantial change of dynamical properties of LDM. Indeed,
let us consider an adiabatically bent ZML. For a conventional Schr\"odinger particle the bending implies a geometric potential, proportional to the square of the curvature at every point \cite{jen71,duc95}. This barrier results in the appearance of a trapped mode and backscattering. Both effects are forbidden for LDM due to the topological protection. In the Section \ref{bent} we show that the bending of ZML, indeed, tends only to the emergence of
geometric phases.

\section{\label{sec::single}Linear dispersion modes and zero-mass lines}

We assume that the dynamics of charge carriers is governed by the Hamiltonian
\begin{equation}
 \hat H=\sigma_x \hat p_x+\sigma_y \hat p_y + \sigma_z m(y),
\label{H}
\end{equation}
where the mass $m(y)$ vanishes along the line $y=0$. Here we put $\hbar=v=1$, where $v$ is the Fermi velocity.
For a given energy $E$ the wavefunction $\Psi$ obeys the equation $\hat H\Psi=E\Psi$.
We write $\Psi=e^{ip_x x}\chi(y)$,
\begin{equation}
 [\sigma_x p_x+\sigma_y \hat p_y + \sigma_z m(y)]\chi(y)=E\chi(y).
\label{HE}
\end{equation}
In a matrix form (\ref{HE}) reads
\begin{equation}
 \left(\begin{array}{cc}m-E & p_x-\partial_y\\p_x+\partial_y & -m-E\end{array}\right)
 \left(\begin{array}{c}\chi_1\\\chi_2\end{array}\right)=0.
\label{mHE}
\end{equation}
Let us now sum up equations in (\ref{mHE}) and subtract the first one from the second. Then we find
\begin{align}
 \left(\begin{array}{cc} p_x-E & \partial_y+m \\
        \partial_y-m & p_x+E\end{array}\right)
        \left(\begin{array}{c}\eta_1 \\ \eta_2\end{array}\right)=0,
\label{HE2}
\end{align}
where we introduced the notations $\eta_1=(\chi_1+\chi_2)/\sqrt{2}$, $\eta_2=(\chi_1-\chi_2)/\sqrt{2}$ (see Appendix \ref{h2} for details).
We can reduce (\ref{HE2}) to scalar Schr\"odinger
equations \cite{volk86}
\begin{align}
 [-\partial_y^2+m(y)^2+m'(y)]\eta_1&=\lambda\eta_1, \label{a1}\\{}
 [-\partial_y^2+m(y)^2-m'(y)]\eta_2&=\lambda\eta_2, \label{a2}
\end{align}
where $E^2=p_x^2+\lambda$. Functions $\eta_1$ and $\eta_2$ are not independent. Connection formulas read
\begin{align}
 (E+p_x)\eta_2=(m-\partial_y)\eta_1,\label{cf}\\
 (E-p_x)\eta_1=(m+\partial_y)\eta_2.\nonumber
\end{align}
Equations (\ref{a1})-(\ref{a2}) can be written in the form
\begin{align}
 (m+\partial_y)(m-\partial_y)\eta_1=\lambda\eta_1,\label{mpmp}\\
 (m-\partial_y)(m+\partial_y)\eta_2=\lambda\eta_2.\nonumber
\end{align}
Multiplying the first equation in (\ref{mpmp}) by $\eta_1$, the second by $\eta_2$ and integrating over $y$ we find
\begin{align}
 \lambda\|\eta_1\|^2=\|(m-\partial_y)\eta_1\|^2\geq 0,\label{etanorm2}\\
 \lambda\|\eta_2\|^2=\|(m+\partial_y)\eta_2\|^2\geq 0,\nonumber
\end{align}
whence $\lambda\geq 0$, since $\eta_1$ and $\eta_2$ can not vanish simultaneously. In (\ref{etanorm2}) we denoted
\begin{equation}
 \|\eta_i\|^2=\int_{-\infty}^\infty \eta_i^2(y)dy.
\end{equation}

Let us consider a case when $m$ monotonously depends on $y$, say $m'(y)>0$. Then equations (\ref{HE2}) comprise a LDM,
for which $E=-p_x$. It is clear that $\eta_1$ should be identically zero, since (\ref{a1}) can not have a zero eigenvalue if the potential is always positive.
From (\ref{HE2}) for $\eta_1=0$ we have $E=-p_x$ and
\begin{equation}
 \eta_2(y)=\exp\left(-\int_0^y dy' m(y')\right).
\label{eta2exact}
\end{equation}

\section{\label{zz}Two zero-mass lines}

Let us consider a mass distribution
\begin{equation}
m(y)=y^2-a^2.
\label{m2}
\end{equation}
It mimics a sample with two parallel ZMLs at $y=\pm a$. The infinitely large values $m(y)$
in the limit $y\to\pm\infty$ can be treated as two edges of the sample.
\begin{figure}
\begin{center}
\includegraphics[width=7cm]{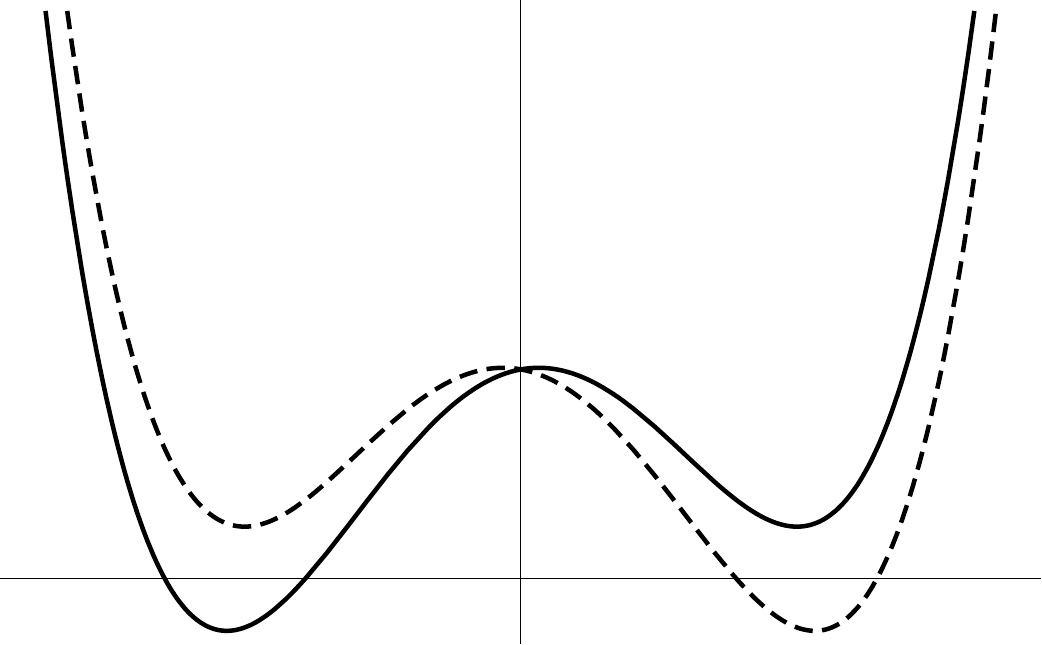}
 \caption{\label{fig::tilted}Tilted double well potentials. Solid and dashed lines correspond to $v_1$, $v_2$ respectively.
 An existence of the upturned well can lead to the Arago effect: focusing in a shadow region}
\end{center}
\end{figure}
Effective potentials
\begin{align}
v_{1}(y)&=m(y)^2+m'(y)=(y^2-a^2)^2+2y,\label{v1}\\
v_{2}(y)&=m(y)^2-m'(y)=(y^2-a^2)^2-2y,\label{v2}
\end{align}
entering equations (\ref{a1}), (\ref{a2}), in this case correspond to tilted double wells (see Fig.~\ref{fig::tilted}).

It is easy to see that $\lambda=0$ does not belong to the spectrum of Eq.~(\ref{a2}). Indeed, the exact solution (\ref{eta2exact}) exponentially decays when $y\to\infty$ and
exponentially grows when $y\to-\infty$. Due to the conservation of Wronskian another linear independent solution of (\ref{a2}) exponentially grows when $y\to\infty$ and exponentially decays when $y\to-\infty$. Similarly we prove that $\lambda=0$ does not belong to the spectrum of Eq. (\ref{a1}).

\begin{figure}
\begin{center}
\includegraphics[width=7cm]{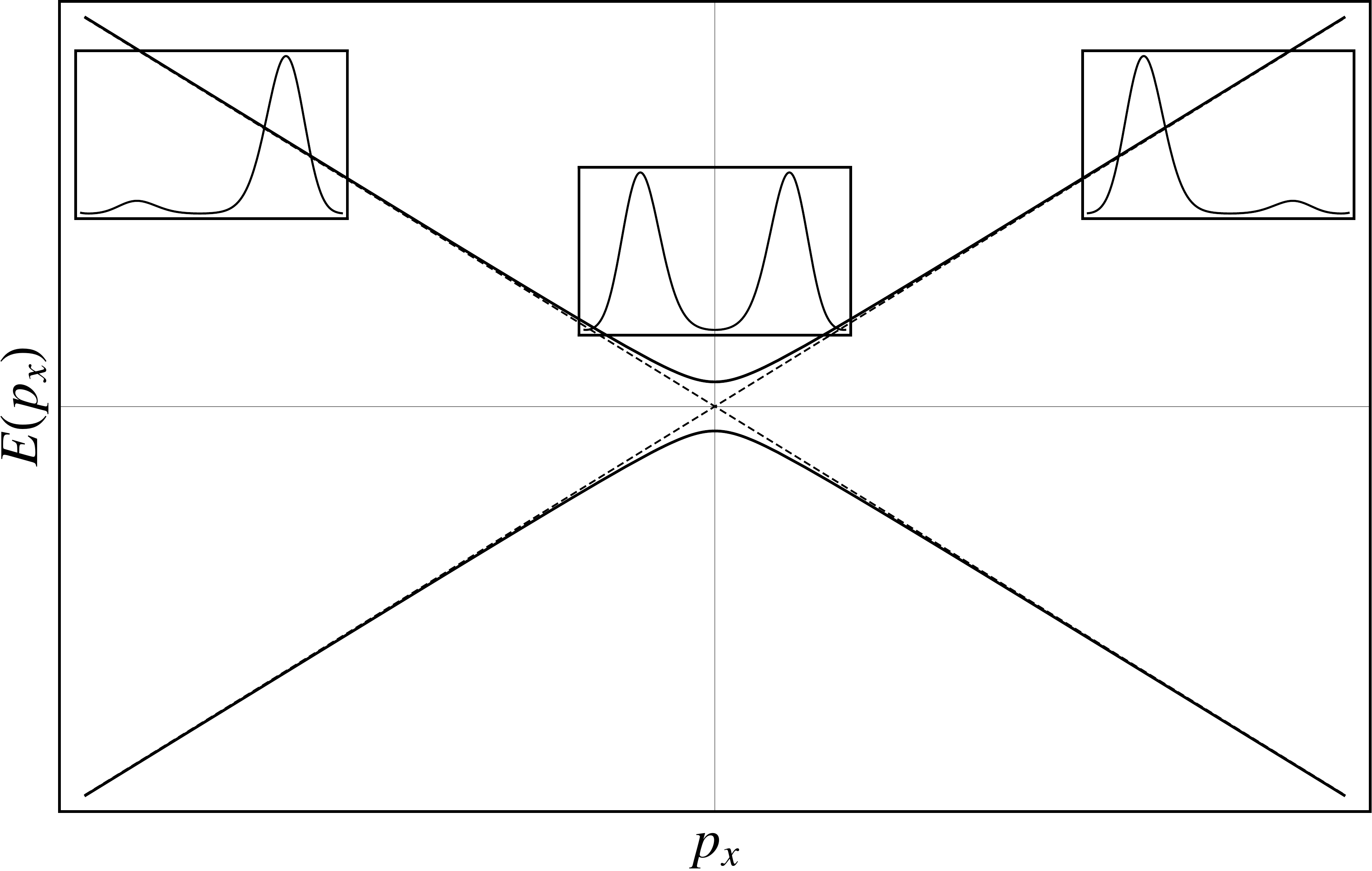}
 \caption{\label{fig::disprel}The schematic view of the dispersion relation and the corresponding wavefunction reconstruction (insets). One sees that at zero
 longitudinal momenta the wavefunction is not localized at a single ZML.}
\end{center}
\end{figure}

Nonzero $\lambda$ leads to avoided crossing of branches $E=\pm p_x$, corresponding to well separated ZMLs (see Fig.~\ref{fig::disprel}). The dispersion relation reads $E_n^\pm(p_x)=\pm\sqrt{p_x^2+\lambda_n}$. The splitting between branches $E_0^\pm(p_x)$ is equal to $2\sqrt{\lambda_0}$, where $\lambda_0$ is the lowest eigenvalue
of the Schr\"odinger equations (\ref{a1}), (\ref{a2}).  We put $p_x=0$ and take the branch corresponding to $E_0(0)=p_y=\sqrt{\lambda_0}$. Equations (\ref{cf}) give
\begin{align}
(m+\partial_y)\tilde\eta_2(y)=p_y\tilde\eta_1(y),\label{t1t2}\\
(m-\partial_y)\tilde\eta_1(y)=p_y\tilde\eta_2(y),\nonumber
\end{align}
where $\tilde\eta_1(y)$, $\tilde\eta_2(y)$ correspond to $p_x=0$. If $\tilde\eta_1(y)$, $\tilde\eta_2(y)$ are known
then the functions $\eta_1(y)$, $\eta_2(y)$ for $p_x\neq 0$ can be reconstructed as
\begin{equation}
\eta_1(y)=(E+p_x)\tilde\eta_1(y), \qquad \eta_2(y)=p_y\tilde\eta_2(y).
\end{equation}
Let us multiply the first equation in (\ref{t1t2}) by $\tilde\eta_1(y)$, the second one
by $\tilde\eta_2(y)$ and subtract the second result from the first one. We obtain
$\partial_y[\tilde\eta_1(y)\tilde\eta_2(y)]=p_y[\tilde\eta_1^2(y)-\tilde\eta_2^2(y)]$.
Integrating the last equality over $y$ from minus to plus infinity we find
$\|\tilde\eta_1\|^2=\|\tilde\eta_2\|^2,$ which suggests that
$\chi_{1,2}(y)=[\tilde\eta_1(y)\pm\tilde\eta_1(y)]/\sqrt{2}$ are not localized at a certain ZML at zero
longitudinal momentum $p_x$, but rather have comparable amplitudes at both ZMLs.
In contrast to the double well problem (see \cite{lanlif3}, p.~183) this effect is not a consequence of a spatial symmetry.
The delocalization at zero longitudinal momentum may destroy the topological
protection against disorder.

Let us now look how the localization appears at non-zero longitudinal momenta. For $|p_x|\gg p_y$ we find $E+p_x\gg p_y$ if $p_x>0$ and
$E+p_x\ll p_y$ if $p_x<0$. Thus for large positive longitudinal momenta $\chi_{1,2}$ are localized at one ZML
and for large negative longitudinal momenta these functions are localized at another ZML.

For symmetric mass distribution (\ref{m2}) we can obtain an analytic estimation for the spitting.
Let us change $y$ to $-y$ in (\ref{t1t2}). Taking into account that $m(y)$ is an even function we find
\begin{align*}
 (m-\partial_y)\tilde\eta_2(-y)=p_y \tilde\eta_1(-y),\\
 (m+\partial_y)\tilde\eta_1(-y)=p_y\tilde\eta_2(-y).
\end{align*}
Thus we conclude $\tilde\eta_1(-y)=\tilde\eta_2(y)$. Let us multiply the first equation in (\ref{t1t2}) by $\tilde\eta_1(y)$ and integrate over $y$. We obtain
\begin{eqnarray}
 p_y\|\tilde\eta_1\|^2&=&\int_{-\infty}^\infty dy\,\tilde\eta_2(-y)(m+\partial_y)\tilde\eta_2(y)=\nonumber\\
 &=&2 \int_0^\infty dy\,\tilde\eta_2(-y)(m+\partial_y)\tilde\eta_2(y)+\tilde\eta_2^2(0)\nonumber\\
 {}&=&2 p_y \int_0^\infty dy\,\tilde\eta_1^2(y)+\tilde\eta_2^2(0),
 \label{doublewell}
\end{eqnarray}
Since the function $\tilde\eta_1$ obeys the Schr\"odinger equation (\ref{a1}) with potential (\ref{v1}) for small $p_y$ it remains exponentially small
for any $y>0$ except the vicinity of the point $y=a$. In this vicinity $\tilde\eta_1$ grows due to the focusing in a shadow region (an effect similar to
the Arago spot, Fig.\ref{fig::tilted}), but nevertheless remains small. Therefore we can neglect an integral in the last equality in (\ref{doublewell}) and write
\begin{equation}
 p_y\|\tilde\eta_2\|^2\simeq \tilde\eta_2^2(0),
\end{equation}
where (\ref{eta2exact}) was used to approximate $\tilde\eta_2$ in the region $y>0$. Using (\ref{eta2exact}) again we find
\begin{equation}
 p_y\int_0^{\infty}dy\,\exp\left(-2\int_0^y dy' m(y')\right)=1.
\label{py}
\end{equation}
The integral in (\ref{py}) can be computed using the Laplace method. It gives
\begin{equation}
 p_y\sqrt{\frac{\pi}{a}}\exp\left(-2\int_0^a dy' m(y')\right)=1,
\end{equation}
whence
\begin{equation}
 p_y=\sqrt{\frac{a}{\pi}}\exp\left(2\int_0^a dy' m(y')\right)=\sqrt{\frac{a}{\pi}}\exp\left(-4a^3/3\right).
\end{equation}
Thus for the splitting we obtain
\begin{equation}
 2|p_y|=2\sqrt{\frac{a}{\pi}}\exp\left(-4a^3/3\right).
\label{splitsym}
\end{equation}

Though the explicit estimation (\ref{splitsym}) of the splitting holds only for the specific symmetric mass distribution (\ref{m2}), one can expect that for a generic case the splitting in LDM for a system with two ZMLs is proportional to
\begin{equation}
 \exp\left(-\int_{a_1}^{a_2}|m(y)|dy \right),
\end{equation}
where $a_1<a_2$ are positions of ZMLs. In contrast to the famous double-well problem, the ground state $\lambda>0$ of (\ref{a1}), (\ref{a2}) is still determined by the tunneling, since for any $m(y)$, every well treated apart generates a zero eigenenergy. For the double well the semiclassical degeneracy persists in the symmetric situation only.

\section{\label{bent}Bent zero-mass line}

Let us consider a bent ZML given by $\{x,y\}=\mathbf{R}(\tau)$, where $\tau$ is a natural parameter, i.e. $|\mathbf{R}'(\tau)|=1$.
In the vicinity of this line we introduce new variables $\tau$, $\xi$ by the equality $\{x,y\}=\mathbf{R}(\tau)+\xi \mathbf{n}(\tau)$, where $\mathbf{n}$ is a unit normal vector on the curve at the point $\tau$. In curvilinear coordinates (\ref{H}) reads
\begin{equation}
 H=-\frac{i\boldsymbol{\sigma}\mathbf{R}'(\tau)}{1-\xi k(\tau)}\frac{\partial}{\partial\tau}
 -i\boldsymbol{\sigma}\mathbf{n}(\tau)\frac{\partial}{\partial\xi}+\sigma_z m,
\end{equation}
where $k(\tau)=-\langle\mathbf{R}',\mathbf{n}'\rangle$ is the curvature at the point $\tau$ and $\boldsymbol{\sigma}\mathbf{b}=\sigma_x b_x+\sigma_y b_y$ for a vector $\mathbf{b}=\{b_x,b_y\}$. Since the Jacobian
\begin{equation}
J=\frac{D(x,y)}{D(\tau,\xi)}=1-k(\tau)\xi
\end{equation}
is not unity, we introduce a new wavefunction
\begin{equation}
\tilde\Psi=\sqrt{1-k(\tau)\xi}\,\Psi,
\end{equation}
which in curvilinear coordinates has a ``conventional'' normalization condition:
\begin{equation}
 \int_V d\tau d\xi (\tilde\Psi^\dagger\tilde\Psi)=1.
\end{equation}
Then the stationary Dirac equation reads $\hat H\tilde\Psi=E\tilde\Psi$,
\begin{align*}
 \hat H&=\frac{\boldsymbol{\sigma}\mathbf{R}'(\tau)}{1-\xi k(\tau)}\hat p_\tau
 -i\boldsymbol{\sigma}\mathbf{n}(\tau)\frac{\partial}{\partial\xi}+\sigma_z m\\
 &-\frac{ik\boldsymbol{\sigma}\mathbf{n}(\tau)}{2(1-\xi k(\tau))}-\frac{i \boldsymbol{\sigma}\mathbf{R}'(\tau)\xi k'(\tau)}{2(1-\xi k(\tau))^2}.
\end{align*}
In the case $|m'_\tau|\ll|m'_\xi|$ we find $\hat H\simeq\hat H_0+\hat H_1$,
\begin{align}
 \hat H_0&=\boldsymbol{\sigma}\mathbf{R}'(\tau)\hat p_\tau
 -i\boldsymbol{\sigma}\mathbf{n}(\tau)\frac{\partial}{\partial\xi}+\sigma_z m, \label{he0}\\
 \hat H_1&=\boldsymbol{\sigma}\mathbf{R}'(\tau)\xi k(\tau)\hat p_\tau
 -\frac{ik}{2}\boldsymbol{\sigma}\mathbf{n}(\tau)\label{he1}.
\end{align}

In the adiabatic approximation the effective dynamics is one-dimensional along ZML. It is governed by the effective scalar Hamiltonian
$\hat L\simeq\hat L_0+\hat L_1$. The symbol \cite{mas81} $L_0(p_\tau,\tau)$ of $\hat L_0$ is an eigenvalue of the problem \cite{bel06}
\begin{align}
 \left(\boldsymbol{\sigma}\mathbf{R}'(\tau) p_\tau
 -i\boldsymbol{\sigma}\mathbf{n}(\tau)\frac{\partial}{\partial\xi}+\sigma_z m\right)\chi(p_\tau,\tau)\nonumber\\
 =L_0(p_\tau,\tau)\chi(p_\tau,\tau).
 \label{L0}
\end{align}
Using the notations $\mathbf{n}=\{n_1,n_2\}$, $\mathbf{R}'=\{n_2,-n_1\}$ we obtain
\begin{equation}
 \left(\begin{array}{cc}m-L_0 & p_\tau-\partial_\xi \\ p_\tau+\partial_\xi & -m-L_0\end{array}\right)
 \left(\begin{array}{c}\tilde\chi_1 \\ \chi_2\end{array}\right)=0,
\label{HE3}
\end{equation}
where $\tilde\chi_1=(n_2-in_1)\chi_1$. After to the replacement $\tau\to x$, $\xi\to y$ (\ref{HE3}) coincides with (\ref{HE2}). From the expression\cite{bel06}
\begin{equation}
 L_1=\langle\chi^\dagger H_1\chi\rangle_\xi+i\left<\chi^\dagger\frac{\partial L_0}{\partial\tau}\frac{\partial\chi}{\partial p_\tau}\right>_\xi
 -i\left<\chi^\dagger\frac{\partial H_0}{\partial p_\tau}\frac{\partial\chi}{\partial \tau}\right>_\xi
\label{L1gen}
\end{equation}
we find
\begin{equation}
 L_1=-\frac{i}{2}\frac{\partial^2 L_0}{\partial p_\tau\partial\tau}
 +\langle\tilde\chi_1\xi\chi_2\rangle_\xi k(\tau)p_\tau
 -\frac{k p_\tau}{2 L_0}.
\label{L1}
\end{equation}
Here $\langle\cdot\rangle_\xi$ means the integration over $\xi$. To pass from (\ref{L1gen}) to (\ref{L1}) we have chosen $\tilde\chi=\{\tilde\chi_1,\chi_2\}$ to be a real function and used the following equalities: $\chi^\dagger(\boldsymbol{\sigma}\mathbf{n})\chi=0$,
$\langle\chi^\dagger \chi'_{p_\tau}\rangle_\xi=0$,
$\chi^\dagger (\boldsymbol{\sigma}\mathbf{R}')\chi=2\tilde\chi_1\chi_2$,
$\chi^\dagger(\boldsymbol{\sigma}\mathbf{R}')\partial_\tau\chi=\left(\partial_\tau-ik\right)\tilde\chi_1\chi_2$,
$\langle\tilde\chi_1\chi_2\rangle_\xi=p_\tau/2L_0$, $(L_0)'_{p_\tau}=p_\tau/L_0$.

The solution $\psi$ of the effective longitudinal equation $\hat L\psi=E\psi$ reads \cite{mas81}
\begin{align}
 \psi(\tau)=\sqrt{\left|\frac{L_0}{p_\tau}\right|}e^{i\theta},\label{psi}\\
 \theta=\int p_\tau d\tau-\int \left(L_0\langle\tilde\chi_1\xi\chi_2\rangle_\xi
 -\frac{1}{2}\right)k(\tau) d\tau.
\end{align}
It relates to the solution of the Dirac equation as $\tilde \Psi=\chi\psi$. From the definition of curvature we have $k(\tau)d\tau=-\langle\mathbf{R}',d\mathbf{n}\rangle$. Introducing the angle $\phi$ between $\mathbf{R}'$ and $x$-axis we find from the last equality $k(\tau)d\tau=d\phi$. This gives
\begin{align}
 \theta=\int p_\tau d\tau-\int L_0\langle\tilde\chi_1\xi\chi_2\rangle_\xi k(\tau) d\tau
 +\frac{\Delta\phi}{2},
\end{align}
where $\Delta\phi$ is the total rotation of the tangent vector.

\begin{figure}
\begin{center}
\includegraphics[width=7cm]{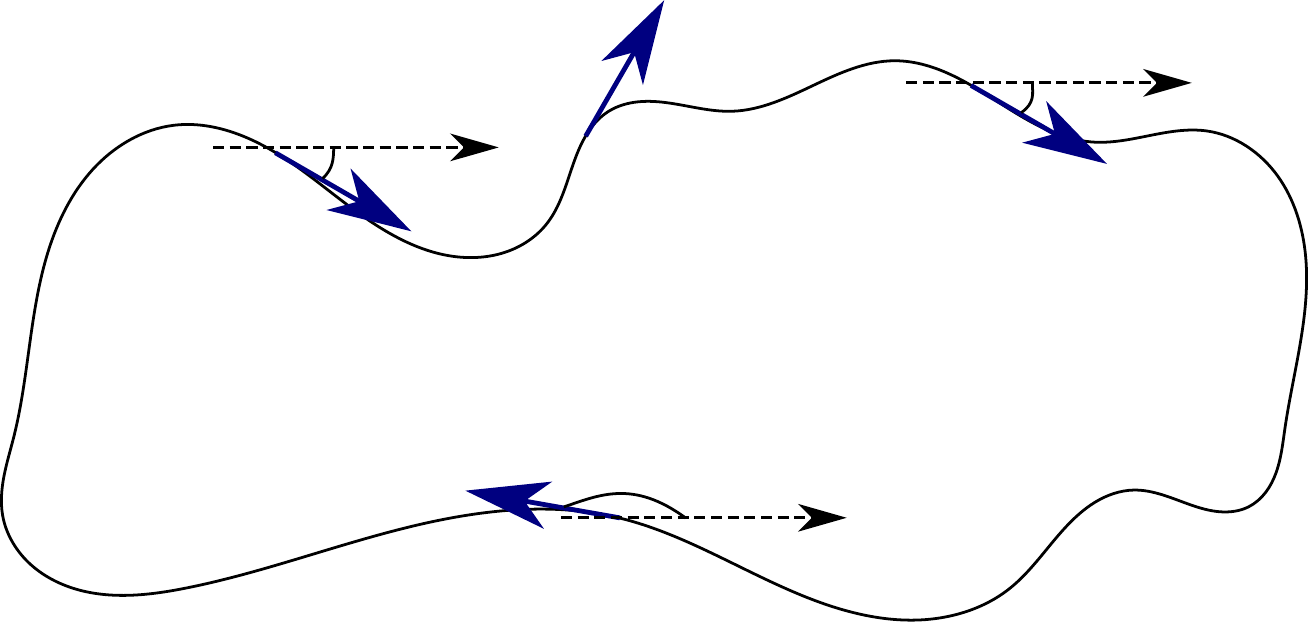}
 \caption{\label{fig::closed} Illustration of the closed ZML. Solid arrows show the tangent vectors, dashed arrows correspond to x-axis.}
\end{center}
\end{figure}

From (\ref{psi}) it is easy to find the quantization rule. For a closed ZML without turning points (see Fig.~\ref{fig::closed}) we obtain
\begin{equation}
\frac{1}{2\pi} \oint p_\tau d\tau
-\frac{E}{2\pi}\oint \langle\tilde\chi_1\xi\chi_2\rangle_\xi k(\tau)d\tau
=n-\frac{w}{2},
\label{q1}
\end{equation}
where $p_\tau=\pm\sqrt{E^2-\lambda(\tau)}$, $\lambda(\tau)=L_0^2(p_\tau=0,\tau)$ and $w$ is the winding number indicating how many times the tangent vector turns around a fixed point. For a curve without intersections $w=1$.

For LDM $L_0=E=-p_\tau$ provided that $m'_\xi>0$ and (\ref{L1}) gives
\begin{equation}
 L_1=\langle\tilde\chi_1\xi\chi_2\rangle_\xi k(\tau)p_\tau+\frac{k}{2}.
\label{eq::L1-zero}
\end{equation}

The first term in (\ref{eq::L1-zero}) describes the energy gain due to the displacement of the Dirac fermion from ZML
caused by the confinement asymmetry. This energy gain is the consequence of the centrifugal force. One sees that in
the considered approximation the centrifugal force itself does not shift the particle since its energy is assumed to be relatively small.
For a higher energy one can use the approach developed in \cite{mas94}.

The term $k/2$ in (\ref{eq::L1-zero}) is a geometric potential associated with a curvature.
It is well-known \cite{jen71,duc95} that for a Schr\"odinger particle in a bent waveguide the geometric potential is negative and proportional to $k^2$. It always leads to the formation of curvature induced bound states. On the contrary, for LDM the geometric potential results in the appearance of a geometric phase only. Indeed, the effective longitudinal wavefunction $\psi$ (\ref{psi}) reads
\begin{eqnarray}
 \psi(\tau)=e^{i\theta},\nonumber\\
 \theta=-E\tau-E\int\langle\tilde\chi_1\xi\chi_2\rangle_\xi k(\tau)d\tau
 +\frac{\Delta\phi}{2}.
 \label{psi0}
\end{eqnarray}
The absence of trapped states for the Dirac fermion can be seen as a manifistation of the Klein tunneling \cite{kat06,tud12} for the massless LDM. A similar effect was already found in \cite{tud06} for a mode with a liner dispersion in a bent graphene ribbon.

For the LDM quantization condition (\ref{q1}) can be simplified to give
\begin{equation}
 -\frac{El}{2\pi}
 -\frac{E}{2\pi}\oint\langle\tilde\chi_1\xi\chi_2\rangle_\xi k(\tau) d\tau=n-\frac{w}{2}.
\label{q10}
\end{equation}
Here $l$ is the length of the closed ZML. From (\ref{q10}) we obtain the semiclassical spectral series $E_n$:
\begin{equation}
 E_n=-\frac{2\pi}{l}\left(n-\frac{w}{2}\right)
 +\frac{2\pi n}{l^2}\oint\langle\tilde\chi_1\xi\chi_2\rangle_\xi
 k(\tau) d\tau.
 \label{eq::L0quant}
\end{equation}
Half-integer numbers in the first term of (\ref{eq::L0quant}) can be seen as a manifestation of the Berry phase for the massless Dirac fermion. Indeed, along ZML a particle described by LDM remains massless, therefore the conventional argument that such a particle acquires the Berry phase equal to $\pi w$ can be repeated.
 
Using the results of Section \ref{sec::single} for a single ZML we find $\chi_1=-\chi_2=\eta_2/\sqrt{2}$, where $\eta_2$ differs from (\ref{eta2exact})
by a normalization factor
\begin{equation}
 N(\tau)=\left[\int_{-\infty}^\infty \exp\left(-2\int_0^\xi d\xi' m(\tau,\xi')\right) d\xi\right]^{-1/2}.
\label{eq::N}
\end{equation}
This gives
\begin{equation}
 \langle\tilde\chi_1\xi\chi_2\rangle_\xi = -\frac{N^2(\tau)}{2}\int_{-\infty}^\infty \xi \exp\left(-2\int_0^\xi d\xi' m(\tau,\xi')\right) d\xi.
 \label{eq::xi}
\end{equation}
More precisely, the integration in (\ref{eq::N}), (\ref{eq::xi}) should be performed between finite limits lying in a sufficiently large vicinity of ZML which, on the other side, does not contain any other ZML. We completely neglected tunneling effects, thus the constructed LDM should be considered as an asymptotic of the eigenfunction of the Dirac equation in the given vicinity of ZML. The non-zero value of (\ref{eq::xi}) means that the Dirac particle is slightly shifted from ZML due to the local asymmetry of the confinement. Indeed, if $m$ is an odd function of $\xi$ the expression (\ref{eq::xi}) vanishes. This happens since the confining potential $m^2(\tau,\xi)$ in this case is symmetric with respect to ZML.

\section{Conclusion}

In the paper we studied the LDM dynamics of charge carriers in topological insulators. We have shown that the wavefunction of a charge carrier is localized along a single ZML only at large longitudinal momenta. At small longitudinal momenta the wavefunction has comparable amplitudes at both edges of the sample, which may affect the topological protection. We found that the curvature of a bent ZML forms a geometric potential, which however does not lead to an appearance of trapped modes due to the Klein tunneling.

\begin{acknowledgements}
We thank Misha Titov, Jan Kees Maan, Uli Zeitler, Laurens Molenkamp, Liv Hornekaer and 
Carlo Beenakker for fruitful discussions.
This work is supported by the Dutch Science Foundation NWO/FOM and the EU-India FP-7 collaboration under MONAMI.
\end{acknowledgements}

\appendix

\section{\label{h2}On the effective Schr\"odinger equation}

In this abstract we would like to clarify the origin of the transformation performed to pass from (\ref{mHE}) to (\ref{HE2}).
Let us write equation (\ref{HE}) in the form $\boldsymbol{\sigma}\hat{\boldsymbol{\pi}}\chi(y)=E\chi(y)$, where $\hat{\boldsymbol{\pi}}=\{p_x,\hat p_y, m(y)\}$.
We can square this equation using the equality
\begin{equation}
 (\boldsymbol{\sigma}\hat{\boldsymbol{\pi}})^2=\hat{\boldsymbol{\pi}}^2+i\sigma_x[\pi_y,\pi_z]+i\sigma_y[\pi_z,\pi_x]+i\sigma_z[\pi_x,\pi_y]
\end{equation}
valid for any non-commuting operators $\hat\pi_x$, $\hat\pi_y$ and $\hat\pi_z$. We find
\begin{equation}
 [p_x^2+\hat p_y^2+m(y)^2+\sigma_x m'(y)]\chi=E^2\chi.
\end{equation}
The last equation comprises the only matrix $\sigma_x$, which does not depend on $y$. We write
\begin{equation}
 \chi=\frac{1}{\sqrt{2}}\left(\begin{array}{c}1 \\ 1\end{array}\right)\eta_1+\frac{1}{\sqrt{2}}\left(\begin{array}{c}1 \\ -1\end{array}\right)\eta_2
\end{equation}
and obtain (\ref{a1}), (\ref{a2}).


\begin{thebibliography}{36}
\expandafter\ifx\csname natexlab\endcsname\relax\def\natexlab#1{#1}\fi
\expandafter\ifx\csname bibnamefont\endcsname\relax
  \def\bibnamefont#1{#1}\fi
\expandafter\ifx\csname bibfnamefont\endcsname\relax
  \def\bibfnamefont#1{#1}\fi
\expandafter\ifx\csname citenamefont\endcsname\relax
  \def\citenamefont#1{#1}\fi
\expandafter\ifx\csname url\endcsname\relax
  \def\url#1{\texttt{#1}}\fi
\expandafter\ifx\csname urlprefix\endcsname\relax\def\urlprefix{URL }\fi
\providecommand{\bibinfo}[2]{#2}
\providecommand{\eprint}[2][]{\url{#2}}

\bibitem[{\citenamefont{Krive and Rozhavskii}(1987)}]{kri87}
\bibinfo{author}{\bibfnamefont{I.}~\bibnamefont{Krive}} \bibnamefont{and}
  \bibinfo{author}{\bibfnamefont{A.}~\bibnamefont{Rozhavskii}},
  \bibinfo{journal}{Sov. Phys. Usp.} \textbf{\bibinfo{volume}{30}},
  \bibinfo{pages}{370} (\bibinfo{year}{1987}).

\bibitem[{\citenamefont{Witten}(1981)}]{wit81}
\bibinfo{author}{\bibfnamefont{E.}~\bibnamefont{Witten}},
  \bibinfo{journal}{Nucl. Phys. B} \textbf{\bibinfo{volume}{188}},
  \bibinfo{pages}{513} (\bibinfo{year}{1981}).

\bibitem[{\citenamefont{Jackiw and Rebbi}(1976)}]{jac76}
\bibinfo{author}{\bibfnamefont{R.}~\bibnamefont{Jackiw}} \bibnamefont{and}
  \bibinfo{author}{\bibfnamefont{C.}~\bibnamefont{Rebbi}},
  \bibinfo{journal}{Phys. Rev. D} \textbf{\bibinfo{volume}{13}},
  \bibinfo{pages}{3398} (\bibinfo{year}{1976}).

\bibitem[{\citenamefont{Su et~al.}(1979)\citenamefont{Su, Schrieffer, and
  Heeger}}]{su79}
\bibinfo{author}{\bibfnamefont{W.~P.} \bibnamefont{Su}},
  \bibinfo{author}{\bibfnamefont{J.~R.} \bibnamefont{Schrieffer}},
  \bibnamefont{and} \bibinfo{author}{\bibfnamefont{A.~J.}
  \bibnamefont{Heeger}}, \bibinfo{journal}{Phys. Rev. Lett.}
  \textbf{\bibinfo{volume}{42}}, \bibinfo{pages}{1698} (\bibinfo{year}{1979}).

\bibitem[{\citenamefont{Volkov and Pankratov}(1986)}]{volk86}
\bibinfo{author}{\bibfnamefont{B.~A.} \bibnamefont{Volkov}} \bibnamefont{and}
  \bibinfo{author}{\bibfnamefont{O.~A.} \bibnamefont{Pankratov}},
  \bibinfo{journal}{Pis'ma Zh. Eksp. Teor. Fiz.} \textbf{\bibinfo{volume}{43}},
  \bibinfo{pages}{99} (\bibinfo{year}{1986}).

\bibitem[{\citenamefont{Ludwig et~al.}(1994)\citenamefont{Ludwig, Fisher,
  Shankar, and Grinstein}}]{lud94}
\bibinfo{author}{\bibfnamefont{A.~W.~W.} \bibnamefont{Ludwig}},
  \bibinfo{author}{\bibfnamefont{M.~P.~A.} \bibnamefont{Fisher}},
  \bibinfo{author}{\bibfnamefont{R.}~\bibnamefont{Shankar}}, \bibnamefont{and}
  \bibinfo{author}{\bibfnamefont{G.}~\bibnamefont{Grinstein}},
  \bibinfo{journal}{Phys. Rev. B} \textbf{\bibinfo{volume}{50}},
  \bibinfo{pages}{7526} (\bibinfo{year}{1994}).

\bibitem[{\citenamefont{Chklovskii et~al.}(1992)\citenamefont{Chklovskii,
  Shklovskii, and Glazman}}]{chk92}
\bibinfo{author}{\bibfnamefont{D.~B.} \bibnamefont{Chklovskii}},
  \bibinfo{author}{\bibfnamefont{B.~I.} \bibnamefont{Shklovskii}},
  \bibnamefont{and} \bibinfo{author}{\bibfnamefont{L.~I.}
  \bibnamefont{Glazman}}, \bibinfo{journal}{Phys. Rev. B}
  \textbf{\bibinfo{volume}{46}}, \bibinfo{pages}{4026} (\bibinfo{year}{1992}).

\bibitem[{\citenamefont{Hettmansperger
  et~al.}(2012)\citenamefont{Hettmansperger, Duerr, Oostinga, Gould,
  Trauzettel, and Molenkamp}}]{het12}
\bibinfo{author}{\bibfnamefont{H.}~\bibnamefont{Hettmansperger}},
  \bibinfo{author}{\bibfnamefont{F.}~\bibnamefont{Duerr}},
  \bibinfo{author}{\bibfnamefont{J.}~\bibnamefont{Oostinga}},
  \bibinfo{author}{\bibfnamefont{C.}~\bibnamefont{Gould}},
  \bibinfo{author}{\bibfnamefont{B.}~\bibnamefont{Trauzettel}},
  \bibnamefont{and}
  \bibinfo{author}{\bibfnamefont{L.}~\bibnamefont{Molenkamp}},
  \bibinfo{journal}{arXiv:1205.5144}  (\bibinfo{year}{2012}).

\bibitem[{\citenamefont{K{\"o}nig et~al.}(2007)\citenamefont{K{\"o}nig,
  Wiedmann, Br{\"u}ne, Roth, Buhmann, Molenkamp, Qi, and Zhang}}]{koe07}
\bibinfo{author}{\bibfnamefont{M.}~\bibnamefont{K{\"o}nig}},
  \bibinfo{author}{\bibfnamefont{S.}~\bibnamefont{Wiedmann}},
  \bibinfo{author}{\bibfnamefont{C.}~\bibnamefont{Br{\"u}ne}},
  \bibinfo{author}{\bibfnamefont{A.}~\bibnamefont{Roth}},
  \bibinfo{author}{\bibfnamefont{H.}~\bibnamefont{Buhmann}},
  \bibinfo{author}{\bibfnamefont{L.}~\bibnamefont{Molenkamp}},
  \bibinfo{author}{\bibfnamefont{X.-L.} \bibnamefont{Qi}}, \bibnamefont{and}
  \bibinfo{author}{\bibfnamefont{S.-C.} \bibnamefont{Zhang}},
  \bibinfo{journal}{Science} \textbf{\bibinfo{volume}{318}},
  \bibinfo{pages}{766} (\bibinfo{year}{2007}).

\bibitem[{\citenamefont{Moore}(2009)}]{M09}
\bibinfo{author}{\bibfnamefont{J.}~\bibnamefont{Moore}},
  \bibinfo{journal}{Nature Physics} \textbf{\bibinfo{volume}{5}},
  \bibinfo{pages}{378} (\bibinfo{year}{2009}).

\bibitem[{\citenamefont{Hasan and Kane}(2010)}]{HK10}
\bibinfo{author}{\bibfnamefont{M.~Z.} \bibnamefont{Hasan}} \bibnamefont{and}
  \bibinfo{author}{\bibfnamefont{C.~L.} \bibnamefont{Kane}},
  \bibinfo{journal}{Rev. Mod. Phys.} \textbf{\bibinfo{volume}{82}},
  \bibinfo{pages}{3045} (\bibinfo{year}{2010}).

\bibitem[{\citenamefont{B{\"u}ttner et~al.}(2011)\citenamefont{B{\"u}ttner,
  Liu, Tkachov, Novik, Br{\"u}ne, Buhmann, Hankiewicz, Recher, Trauzettel,
  Zhang et~al.}}]{bue11}
\bibinfo{author}{\bibfnamefont{B.}~\bibnamefont{B{\"u}ttner}},
  \bibinfo{author}{\bibfnamefont{C.~X.} \bibnamefont{Liu}},
  \bibinfo{author}{\bibfnamefont{G.}~\bibnamefont{Tkachov}},
  \bibinfo{author}{\bibfnamefont{E.~G.} \bibnamefont{Novik}},
  \bibinfo{author}{\bibfnamefont{C.}~\bibnamefont{Br{\"u}ne}},
  \bibinfo{author}{\bibfnamefont{H.}~\bibnamefont{Buhmann}},
  \bibinfo{author}{\bibfnamefont{E.~M.} \bibnamefont{Hankiewicz}},
  \bibinfo{author}{\bibfnamefont{P.}~\bibnamefont{Recher}},
  \bibinfo{author}{\bibfnamefont{B.}~\bibnamefont{Trauzettel}},
  \bibinfo{author}{\bibfnamefont{S.~C.} \bibnamefont{Zhang}},
  \bibnamefont{et~al.}, \bibinfo{journal}{Nature Physics}
  \textbf{\bibinfo{volume}{7}}, \bibinfo{pages}{418} (\bibinfo{year}{2011}).

\bibitem[{\citenamefont{Qi and Zhang}(2011)}]{qi11}
\bibinfo{author}{\bibfnamefont{X.-L.} \bibnamefont{Qi}} \bibnamefont{and}
  \bibinfo{author}{\bibfnamefont{S.-C.} \bibnamefont{Zhang}},
  \bibinfo{journal}{Rev. Mod. Phys.} \textbf{\bibinfo{volume}{83}},
  \bibinfo{pages}{1057} (\bibinfo{year}{2011}).

\bibitem[{\citenamefont{Sachs et~al.}(2011)\citenamefont{Sachs, Wehling,
  Katsnelson, and Lichtenstein}}]{sac11}
\bibinfo{author}{\bibfnamefont{B.}~\bibnamefont{Sachs}},
  \bibinfo{author}{\bibfnamefont{T.~O.} \bibnamefont{Wehling}},
  \bibinfo{author}{\bibfnamefont{M.~I.} \bibnamefont{Katsnelson}},
  \bibnamefont{and} \bibinfo{author}{\bibfnamefont{A.~I.}
  \bibnamefont{Lichtenstein}}, \bibinfo{journal}{Phys. Rev. B}
  \textbf{\bibinfo{volume}{84}}, \bibinfo{pages}{195414}
  (\bibinfo{year}{2011}).

\bibitem[{\citenamefont{Zarenia et~al.}(2012)\citenamefont{Zarenia, Leenaerts,
  Partoens, and Peeters}}]{zar12}
\bibinfo{author}{\bibfnamefont{M.}~\bibnamefont{Zarenia}},
  \bibinfo{author}{\bibfnamefont{O.}~\bibnamefont{Leenaerts}},
  \bibinfo{author}{\bibfnamefont{B.}~\bibnamefont{Partoens}}, \bibnamefont{and}
  \bibinfo{author}{\bibfnamefont{F.~M.} \bibnamefont{Peeters}},
  \bibinfo{journal}{to be published}  (\bibinfo{year}{2012}).

\bibitem[{\citenamefont{Elias et~al.}(2009)\citenamefont{Elias, Nair,
  Mohiuddin, Morozov, Blake, Halsall, Ferrari, Boukhvalov, Katsnelson, Geim
  et~al.}}]{eli09}
\bibinfo{author}{\bibfnamefont{D.~C.} \bibnamefont{Elias}},
  \bibinfo{author}{\bibfnamefont{R.~R.} \bibnamefont{Nair}},
  \bibinfo{author}{\bibfnamefont{T.~M.~G.} \bibnamefont{Mohiuddin}},
  \bibinfo{author}{\bibfnamefont{S.~V.} \bibnamefont{Morozov}},
  \bibinfo{author}{\bibfnamefont{P.}~\bibnamefont{Blake}},
  \bibinfo{author}{\bibfnamefont{M.~P.} \bibnamefont{Halsall}},
  \bibinfo{author}{\bibfnamefont{A.~C.} \bibnamefont{Ferrari}},
  \bibinfo{author}{\bibfnamefont{D.~W.} \bibnamefont{Boukhvalov}},
  \bibinfo{author}{\bibfnamefont{M.~I.} \bibnamefont{Katsnelson}},
  \bibinfo{author}{\bibfnamefont{A.~K.} \bibnamefont{Geim}},
  \bibnamefont{et~al.}, \bibinfo{journal}{Science}
  \textbf{\bibinfo{volume}{323}}, \bibinfo{pages}{610} (\bibinfo{year}{2009}).

\bibitem[{\citenamefont{Balog et~al.}(2010)\citenamefont{Balog, J{\o}rgensen,
  Nilsson, Andersen, Rienks, Bianchi, Fanetti, Laegsgaard, Baraldi, Lizzit
  et~al.}}]{bal10}
\bibinfo{author}{\bibfnamefont{R.}~\bibnamefont{Balog}},
  \bibinfo{author}{\bibfnamefont{B.}~\bibnamefont{J{\o}rgensen}},
  \bibinfo{author}{\bibfnamefont{L.}~\bibnamefont{Nilsson}},
  \bibinfo{author}{\bibfnamefont{M.}~\bibnamefont{Andersen}},
  \bibinfo{author}{\bibfnamefont{E.}~\bibnamefont{Rienks}},
  \bibinfo{author}{\bibfnamefont{M.}~\bibnamefont{Bianchi}},
  \bibinfo{author}{\bibfnamefont{M.}~\bibnamefont{Fanetti}},
  \bibinfo{author}{\bibfnamefont{E.}~\bibnamefont{Laegsgaard}},
  \bibinfo{author}{\bibfnamefont{A.}~\bibnamefont{Baraldi}},
  \bibinfo{author}{\bibfnamefont{S.}~\bibnamefont{Lizzit}},
  \bibnamefont{et~al.}, \bibinfo{journal}{Nature Materials}
  \textbf{\bibinfo{volume}{9}}, \bibinfo{pages}{315} (\bibinfo{year}{2010}).

\bibitem[{\citenamefont{Nair et~al.}(2010)\citenamefont{Nair, Ren, Jalil, Riaz,
  Kravets, Britnell, Blake, Schedin, Mayorov, Yuan et~al.}}]{rah10}
\bibinfo{author}{\bibfnamefont{R.~R.} \bibnamefont{Nair}},
  \bibinfo{author}{\bibfnamefont{W.}~\bibnamefont{Ren}},
  \bibinfo{author}{\bibfnamefont{R.}~\bibnamefont{Jalil}},
  \bibinfo{author}{\bibfnamefont{I.}~\bibnamefont{Riaz}},
  \bibinfo{author}{\bibfnamefont{V.~G.} \bibnamefont{Kravets}},
  \bibinfo{author}{\bibfnamefont{L.}~\bibnamefont{Britnell}},
  \bibinfo{author}{\bibfnamefont{P.}~\bibnamefont{Blake}},
  \bibinfo{author}{\bibfnamefont{F.}~\bibnamefont{Schedin}},
  \bibinfo{author}{\bibfnamefont{A.~S.} \bibnamefont{Mayorov}},
  \bibinfo{author}{\bibfnamefont{S.}~\bibnamefont{Yuan}}, \bibnamefont{et~al.},
  \bibinfo{journal}{Small} \textbf{\bibinfo{volume}{6}}, \bibinfo{pages}{2877}
  (\bibinfo{year}{2010}).

\bibitem[{\citenamefont{Martin et~al.}(2008)\citenamefont{Martin, Blanter, and
  Morpurgo}}]{mar08}
\bibinfo{author}{\bibfnamefont{I.}~\bibnamefont{Martin}},
  \bibinfo{author}{\bibfnamefont{Y.~M.} \bibnamefont{Blanter}},
  \bibnamefont{and} \bibinfo{author}{\bibfnamefont{A.~F.}
  \bibnamefont{Morpurgo}}, \bibinfo{journal}{Phys. Rev. Lett.}
  \textbf{\bibinfo{volume}{100}}, \bibinfo{pages}{036804}
  (\bibinfo{year}{2008}).

\bibitem[{\citenamefont{Li et~al.}(2011)\citenamefont{Li, Martin, B{\"u}ttiker,
  and Morpurgo}}]{li11}
\bibinfo{author}{\bibfnamefont{J.}~\bibnamefont{Li}},
  \bibinfo{author}{\bibfnamefont{I.}~\bibnamefont{Martin}},
  \bibinfo{author}{\bibfnamefont{M.}~\bibnamefont{B{\"u}ttiker}},
  \bibnamefont{and} \bibinfo{author}{\bibfnamefont{A.~F.}
  \bibnamefont{Morpurgo}}, \bibinfo{journal}{Nature Physics}
  \textbf{\bibinfo{volume}{7}}, \bibinfo{pages}{38} (\bibinfo{year}{2011}).

\bibitem[{\citenamefont{Zarenia et~al.}(2011)\citenamefont{Zarenia, Pereira,
  Farias, and Peeters}}]{zar11}
\bibinfo{author}{\bibfnamefont{M.}~\bibnamefont{Zarenia}},
  \bibinfo{author}{\bibfnamefont{J.~M.} \bibnamefont{Pereira}},
  \bibinfo{author}{\bibfnamefont{G.~A.} \bibnamefont{Farias}},
  \bibnamefont{and} \bibinfo{author}{\bibfnamefont{F.~M.}
  \bibnamefont{Peeters}}, \bibinfo{journal}{Phys. Rev. B}
  \textbf{\bibinfo{volume}{84}}, \bibinfo{pages}{125451}
  (\bibinfo{year}{2011}).

\bibitem[{\citenamefont{Bardarson et~al.}(2010)\citenamefont{Bardarson,
  Medvedyeva, Tworzyd{\l}o, Akhmerov, and Beenakker}}]{bar10}
\bibinfo{author}{\bibfnamefont{J.~H.} \bibnamefont{Bardarson}},
  \bibinfo{author}{\bibfnamefont{M.~V.} \bibnamefont{Medvedyeva}},
  \bibinfo{author}{\bibfnamefont{J.}~\bibnamefont{Tworzyd{\l}o}},
  \bibinfo{author}{\bibfnamefont{A.~R.} \bibnamefont{Akhmerov}},
  \bibnamefont{and} \bibinfo{author}{\bibfnamefont{C.~W.~J.}
  \bibnamefont{Beenakker}}, \bibinfo{journal}{Phys. Rev. B}
  \textbf{\bibinfo{volume}{81}}, \bibinfo{pages}{121414(R)}
  (\bibinfo{year}{2010}).

\bibitem[{\citenamefont{Halperin}(1982)}]{hal82}
\bibinfo{author}{\bibfnamefont{B.~I.} \bibnamefont{Halperin}},
  \bibinfo{journal}{Phys. Rev. B} \textbf{\bibinfo{volume}{25}},
  \bibinfo{pages}{2185} (\bibinfo{year}{1982}).

\bibitem[{\citenamefont{Kane and Mele}(2005)}]{KM05b}
\bibinfo{author}{\bibfnamefont{C.~L.} \bibnamefont{Kane}} \bibnamefont{and}
  \bibinfo{author}{\bibfnamefont{E.~J.} \bibnamefont{Mele}},
  \bibinfo{journal}{Phys. Rev. Lett.} \textbf{\bibinfo{volume}{95}},
  \bibinfo{pages}{146802} (\bibinfo{year}{2005}).

\bibitem[{\citenamefont{Bernevig et~al.}(2006)\citenamefont{Bernevig, Hughes,
  and Zhang}}]{ber06}
\bibinfo{author}{\bibfnamefont{B.~A.} \bibnamefont{Bernevig}},
  \bibinfo{author}{\bibfnamefont{T.~L.} \bibnamefont{Hughes}},
  \bibnamefont{and} \bibinfo{author}{\bibfnamefont{S.-C.} \bibnamefont{Zhang}},
  \bibinfo{journal}{Science} \textbf{\bibinfo{volume}{314}},
  \bibinfo{pages}{1757} (\bibinfo{year}{2006}).

\bibitem[{\citenamefont{Kang et~al.}(2000)\citenamefont{Kang, Stormer,
  Pfeiffer, Baldwin, and West}}]{kan00}
\bibinfo{author}{\bibfnamefont{W.}~\bibnamefont{Kang}},
  \bibinfo{author}{\bibfnamefont{H.~L.} \bibnamefont{Stormer}},
  \bibinfo{author}{\bibfnamefont{L.~N.} \bibnamefont{Pfeiffer}},
  \bibinfo{author}{\bibfnamefont{K.~W.} \bibnamefont{Baldwin}},
  \bibnamefont{and} \bibinfo{author}{\bibfnamefont{K.~W.} \bibnamefont{West}},
  \bibinfo{journal}{Nature} \textbf{\bibinfo{volume}{403}}, \bibinfo{pages}{59}
  (\bibinfo{year}{2000}).

\bibitem[{\citenamefont{Str{\"o}m and Johannesson}(2009)}]{str09}
\bibinfo{author}{\bibfnamefont{A.}~\bibnamefont{Str{\"o}m}} \bibnamefont{and}
  \bibinfo{author}{\bibfnamefont{H.}~\bibnamefont{Johannesson}},
  \bibinfo{journal}{Phys. Rev. Lett.} \textbf{\bibinfo{volume}{102}},
  \bibinfo{pages}{096806} (\bibinfo{year}{2009}).

\bibitem[{\citenamefont{Jensen and Koppe}(1971)}]{jen71}
\bibinfo{author}{\bibfnamefont{H.}~\bibnamefont{Jensen}} \bibnamefont{and}
  \bibinfo{author}{\bibfnamefont{H.}~\bibnamefont{Koppe}},
  \bibinfo{journal}{Annals of Physics} \textbf{\bibinfo{volume}{63}},
  \bibinfo{pages}{586} (\bibinfo{year}{1971}).

\bibitem[{\citenamefont{Duclos and Exner}(1995)}]{duc95}
\bibinfo{author}{\bibfnamefont{P.}~\bibnamefont{Duclos}} \bibnamefont{and}
  \bibinfo{author}{\bibfnamefont{P.}~\bibnamefont{Exner}},
  \bibinfo{journal}{{R}eviews in {M}athematical {P}hysics}
  \textbf{\bibinfo{volume}{7}}, \bibinfo{pages}{73} (\bibinfo{year}{1995}).

\bibitem[{\citenamefont{Landau and Lifshitz}(1977)}]{lanlif3}
\bibinfo{author}{\bibfnamefont{L.~D.} \bibnamefont{Landau}} \bibnamefont{and}
  \bibinfo{author}{\bibfnamefont{E.~M.} \bibnamefont{Lifshitz}},
  \emph{\bibinfo{title}{Quantum Mechanics. Non-relativistic theory}},
  vol.~\bibinfo{volume}{3} (\bibinfo{publisher}{Pergamon Press},
  \bibinfo{year}{1977}).

\bibitem[{\citenamefont{Maslov and Fedoryuk}(1981)}]{mas81}
\bibinfo{author}{\bibfnamefont{V.~P.} \bibnamefont{Maslov}} \bibnamefont{and}
  \bibinfo{author}{\bibfnamefont{M.~V.} \bibnamefont{Fedoryuk}},
  \emph{\bibinfo{title}{Semi-Classical Approximation in Quantum Mechanics}},
  Mathematical Physics and Applied Mathematics 7, Contemporary Mathematics 5
  (\bibinfo{publisher}{D. Reidel Publishing Co.}, \bibinfo{address}{Dordrecht},
  \bibinfo{year}{1981}).

\bibitem[{\citenamefont{Belov et~al.}(2006)\citenamefont{Belov, Dobrokhotov,
  and Tudorovskiy}}]{bel06}
\bibinfo{author}{\bibfnamefont{V.~V.} \bibnamefont{Belov}},
  \bibinfo{author}{\bibfnamefont{S.~Y.} \bibnamefont{Dobrokhotov}},
  \bibnamefont{and} \bibinfo{author}{\bibfnamefont{T.~Y.}
  \bibnamefont{Tudorovskiy}}, \bibinfo{journal}{J. Eng. Math.}
  \textbf{\bibinfo{volume}{55}}, \bibinfo{pages}{179} (\bibinfo{year}{2006}).

\bibitem[{\citenamefont{Maslov}(1994)}]{mas94}
\bibinfo{author}{\bibfnamefont{V.~P.} \bibnamefont{Maslov}},
  \emph{\bibinfo{title}{The Complex WKB Method for Nonlinear Equations I:
  Linear Theory}} (\bibinfo{publisher}{Birkh{\"a}ser Verlag},
  \bibinfo{year}{1994}).

\bibitem[{\citenamefont{Katsnelson et~al.}(2006)\citenamefont{Katsnelson,
  Novoselov, and Geim}}]{kat06}
\bibinfo{author}{\bibfnamefont{M.~I.} \bibnamefont{Katsnelson}},
  \bibinfo{author}{\bibfnamefont{K.~S.} \bibnamefont{Novoselov}},
  \bibnamefont{and} \bibinfo{author}{\bibfnamefont{A.~K.} \bibnamefont{Geim}},
  \bibinfo{journal}{Nature Phys.} \textbf{\bibinfo{volume}{2}},
  \bibinfo{pages}{620} (\bibinfo{year}{2006}).

\bibitem[{\citenamefont{Tudorovskiy et~al.}(2012)\citenamefont{Tudorovskiy,
  Reijnders, and Katsnelson}}]{tud12}
\bibinfo{author}{\bibfnamefont{T.}~\bibnamefont{Tudorovskiy}},
  \bibinfo{author}{\bibfnamefont{K.~J.~A.} \bibnamefont{Reijnders}},
  \bibnamefont{and} \bibinfo{author}{\bibfnamefont{M.~I.}
  \bibnamefont{Katsnelson}}, \bibinfo{journal}{Phys. Scr. T}
  \textbf{\bibinfo{volume}{146}}, \bibinfo{pages}{014010}
  (\bibinfo{year}{2012}).

\bibitem[{\citenamefont{Tudorovskiy and Chaplik}(2006)}]{tud06}
\bibinfo{author}{\bibfnamefont{T.}~\bibnamefont{Tudorovskiy}} \bibnamefont{and}
  \bibinfo{author}{\bibfnamefont{A.~V.} \bibnamefont{Chaplik}},
  \bibinfo{journal}{Pis'ma Zh. Eksp. Teor. Fiz.} \textbf{\bibinfo{volume}{84}},
  \bibinfo{pages}{619} (\bibinfo{year}{2006}).

\end{thebibliography}

\end{document}